# To Study the effects of isospin cross section on mass asymmetric nuclear reactions

Deepinder Kaur*, Varinderjit Kaur and Suneel Kumar
*School of Physics and Materials Science, Thapar University, Patiala – 147004 (Punjab) INDIA.*
*email:deep8894@gmail.com

## Introduction:

The growing research in the experimental isospin physics has motivated us to perform theoretical study regarding the dynamics governing the multifragmentation. Heavy ion collisions offer the possibility to probe nuclear matter under different conditions of densities and temperature. At high excitation energies and temperature, the nuclei may break up into many fragments known as multifragmentation. In the recent times a lot of research is going on for the study of collision of mass asymmetric nuclei at intermediate energy [1]. Multifragmentation is by essence associated to the emission of several fragments. Any study of the phenomenon requires a coincident and efficient detection of these fragments and of the associated particles (Z≤2). In the recent years, multifragmentation studies were performed with $4\pi$ detectors [2].

The isospin effects of the in-medium NN cross section on the physical quantities arise from the difference between isospin dependent in-medium NN cross section denoted by $\sigma_{iso}$ in which $\sigma_{np} > \sigma_{nn} = \sigma_{pp}$ and isospin independent NN cross section denoted by $\sigma_{noiso}$ in which $\sigma_{np} = \sigma_{nn} = \sigma_{pp}$. Here $\sigma_{np}$, $\sigma_{nn}$ and $\sigma_{pp}$ *are* the neutron–proton, neutron–neutron and proton–proton cross sections, respectively.

Jian-Ye Liu et. at. studied the isospin effects of one-body dissipation and two-body collision on the number of protons (neutrons) emitted during the nuclear reaction. Their studies show strongly that the isospin-dependent in-medium NN cross section has a much stronger influence on NP(NN) ( the number of proton(neutron) emissions) [3]. We will see this effect on mass asymmetric systems. The mass asymmetry of a reaction can be defined by the asymmetry parameter $\eta = |(A_T - A_P)/(A_T + A_P)|$; [1] where $A_T$ and $A_P$ are the masses of the target and projectile, respectively. The $\eta = 0$ corresponds to the symmetric reactions, whereas non-zero values of η define different asymmetries of a reaction. Reactions include $^{208}Pb_{82} + ^{208}Pb_{82}$ having $\eta = 0$, $^{40}Ca_{20} + ^{208}Pb_{82}$ having $\eta = 0.6$ and $^{12}C_6 + ^{197}Au_{79}$ having $\eta = 0.8$   This work has been carried out within the frame work of isospin dependent quantum molecular dynamics (IQMD) model.

## The Model:

The dynamical model we used for the study is isospin quantum molecular dynamics (IQMD). The Isospin-dependent Quantum Molecular Dynamic model (IQMD) [4] is the refinement of QMD model based on event by event method [5] The baryons are represented by Gaussian-shaped density distributions

$$f_i(\vec{r},\vec{p},t) = \frac{1}{\pi^2 \hbar^2} e^{-[\vec{r}-\vec{r}_i(t)]^2 \frac{1}{2L}} e^{-[\vec{p}-\vec{p}_i(t)]^2 \frac{2L}{\hbar^2}}$$

The successfully initialized nuclei are then boosted towards each other using Hamilton equations of motion

$$\frac{dr_i}{dt} = \frac{d\langle H \rangle}{d p_i} \; ; \; \frac{dp_i}{dt} = - \frac{d\langle H \rangle}{d r_i}$$

With $\langle H \rangle = \langle T \rangle + \langle V \rangle$ is the total Hamiltonian.

$$f_i(\vec{r},\vec{p},t) = \frac{1}{(\pi\hbar)^3} \times e^{[-(\vec{r}-\vec{r}_i(t))^2 \frac{2}{L}]} \times e^{[-(\vec{p}-\vec{p}_i(t))^2 \frac{L}{2\hbar^2}]}$$

The total potential is the sum of the following specific elementary potentials.

$$V = V_{Sky} + V_{Yuk} + V_{Coul} + V_{mdi} + V_{loc}$$

During the propagation, two nucleons are supposed to suffer a binary collision if the distance between their centroid is

$$|r_i - r_j| \leq \sqrt{\frac{\sigma_{tot}}{\pi}}$$

Where $\sigma_{tot} = \sigma(\sqrt{s}, type)$

The collision is blocked with a possibility

$$P_{block} = 1-(1-P_i)(1-P_j)$$

Where $P_i$ and $P_j$ are the already occupied phase space fractions by other nucleons.

## Results and discussions:

In the figure1 we have displayed the free nucleons and LMF's with isospin dependent as well as isospin independent cross section as a function of energy for symmetric systems and asymmetric systems for scaled impact parameters. As it is clear from the figure that the number of free nucleons is increasing with the increase in energy. This is due to the reason that at central collision all the nucleons are taking part in the collision. The collisions become more violent as the energy is increases. The maximum number of free nucleons will be produced at high energy due to more compression zone. Also with increase in energy Pauli blocking effect decreases. The correlations among the nucleons are destroyed at high energies and hence more number of free nucleons are produced. With the increases in the value of scaled impact parameter the multiplicity of free nucleons is decreasing as compared to central collision. $\sigma_{noiso}$ lead to enhanced production of free nucleons at all the energies and for all asymmetries. For symmetric reactions the difference in the production of free nucleons is more for the energy range 400 MeV/nucleon to 1000 MeV/nucleon. But the asymmetric reactions, the influence is more from 200MeV/nucleon to 600MeV/nucleon.

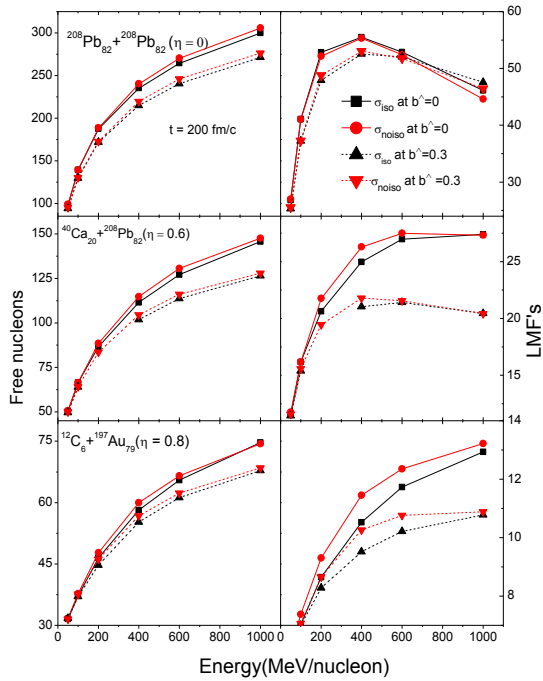

Fig.1: Multiplicity of free nucleons and LMF's as a function of energy.

From the figure, it is clear that the number of LMF's first increases as energy increases, reaches a peak value at 200MeV/nucleon and then decreases as the energy increases. This trend is for symmetric systems i.e for $^{208}Pb_{82}+^{208}Pb_{82}$. But this trend is not followed by mass asymmetric systems. i.e $^{12}C_6+^{197}Au_{79}$ and $^{40}Ca_{20}+^{208}Pb_{82}$. In mass asymmetric systems the number of LMF's increases with the increase in energy. For asymmetric systems the difference is more because mass asymmetry play significant role on reaction dynamics as studied in the ref [1].

Now if effect of $\sigma_{noiso}$ is observed then it is seen that there is considerable effect of cross section on the mass asymmetric systems than on symmetric systems. It can be seen from the figure that the number of light mass fragments formed without isospin dependent nucleon nucleon cross section is more as compared to symmetric systems. In case of symmetric systems the number of LMF's is decreasing with increasing energy by taking the isospin independent cross section.